\documentclass[prb,showpacs,byrevtex,floatfix,twocolumn,superscriptaddress]{revtex4}
\usepackage{amssymb,graphicx,afterpage}
\begin{document}
\title{\boldmath High-pressure infrared spectroscopy: tuning of the low-energy excitations in correlated electron systems \unboldmath}
%
%
\author{I. K\'ezsm\'arki}
\affiliation{Department of Physics, Budapest University of
Technology and Economics, 1111 Budapest, Hungary}
\author{R. Ga\'al}
\affiliation{Ecole Politechnique Federale, Lausanne, CH-1015
Switzerland}
\author{C.C. Homes}
\affiliation{Condensed Matter Physics and Materials Sciences
Department, Brookhaven National Laboratory, Upton, NY 11973 USA}
\author{B. S\'ipos}
\affiliation{Ecole Politechnique Federale, Lausanne, CH-1015
Switzerland}
\author{H. Berger}
\affiliation{Ecole Politechnique Federale, Lausanne, CH-1015
Switzerland}
\author{S. Bord\'acs}
\affiliation{Department of Physics, Budapest University of
Technology and Economics, 1111 Budapest, Hungary}
\author{G. Mih\'aly}
\affiliation{Department of Physics, Budapest University of
Technology and Economics, 1111 Budapest, Hungary}
\author{L. Forr\'o}
\affiliation{Ecole Politechnique Federale, Lausanne, CH-1015
Switzerland}
\date{\today}
%
%
\pacs{\ }
\begin{abstract}
We have extended the range of the high-pressure optical spectroscopy
to the far-infrared region keeping the accuracy of ambient-pressure
experiments. The newly-developed method offers a powerful tool for
the study of pressure-induced phase transitions and
electronic-structural changes in correlated electron systems. The
novel-type optical pressure cell, equipped with large free-aperture
diamond window, allows the measurement of optical reflectivity down
to $\omega\approx20-30$\,cm$^{-1}$ for hydrostatic pressures up to
$p\approx26$\,kbar. The efficiency of the technique is demonstrated
by the investigation of the 2-dimensional charge-density-wave
1$T$-TaS$_2$ whose electronic structure shows high sensitivity to
external pressure. The room-temperature semi-metallic phase of
1$T$-TaS$_2$ is effectively extended by application of pressure and
stabilized as the ground state above $p=14$\,kbar. The corresponding
fully incoherent low-energy optical conductivity is almost
temperature independent below $T=300$\,K. For intermediate
pressures, the onset of the low-temperature insulating phase is
reflected by the sudden drop of the reflectivity and by the
emergence of sharp phonon resonances.
\end{abstract}
\maketitle
%
The application of hydrostatic pressure offers a clean and
controllable way to fine-tune the electronic and magnetic properties
of solids via the change of their bandwidth/dimensionality or by
altering the energy scale of the relevant interactions, such as
magnetic exchange interactions or the electron-phonon coupling. The
possibility to gain spectroscopic information about the electronic
structural changes upon pressure-induced phase transitions has
recently attracted broad interest. As one of the earliest trials,
the possible metallization of solid hydrogen was studied under
extremely high pressures ($p\sim1500$\,kbar) by infrared absorption
and reflectivity techniques.\cite{Mao1990,Hemley1996} In correlated
electron systems, due to the strong competition between neighboring
thermodynamic phases, moderate pressures can effectively induce
phase transitions. Therefore, hydrostatic pressure has been
successfully applied to investigate the phase diagram of correlated
electron materials, especially for compounds located in the vicinity
of a insulator-metal phase boundary. The pressure-driven collapse of
the insulating state has been recently followed by infrared
spectroscopy in the colossal magnetoresistance manganite
La$_{1-x}$Ca$_x$MnO$_3$,\cite{Sacchetti2006} in Mott insulators such
as YNiO$_3$ \cite{Garcia2004} and V$_3$O$_5$,\cite{Baldassarre2007}
in the band insulator YH$_3$,\cite{Ohmura2006} and in the
charge-density-wave rare-earth tritellurides.\cite{Sacchetti2007}
Besides the phase transition phenomena, strong pressure-induced
variation of the electronic structure has been reported for
CeSb,\cite{Nishi2005}
$\beta$-Na$_{0.33}$V$_2$O$_5$,\cite{Kuntscher2005}
silane,\cite{Sun2006} etc. All of these experiments were performed
using diamond anvil cells (DAC) which allowed the generation of
hydrostatic pressure up to $p\approx260$\,kbar. The disadvantage of
DACs is their pressure-limited small aperture, typically
$\sim100$\,$\mu$m in diameter, which introduces a low-energy cutoff
(a few hundred cm$^{-1}$) due to the diffraction effects.
Consequently, these investigations were restricted to the
mid-infrared and higher energy range, although owing to its high
transparency diamond is an ideal optical-window material down to
microwave frequencies.\cite{Edwards1985}

In strongly correlated electron systems, where usually the
low-energy excitations are of interest, spectral information about
the far-infrared region is highly desirable. Focusing on materials
which show high sensitivity to external pressure, we have developed
a conceptually new type of optical pressure cell which extends the
low-energy limit of the infrared experiments by more than one decade
as it is applicable typically down to $\omega\approx
20-30$\,cm$^{-1}$. In our design, this improvement is the result of
the large free aperture of the window ($d=1.5$\,mm) which restricts
the maximum applied pressure to $p\approx26$\,kbar at the same time.
On the basis of reflectivity measurements performed on the two
dimensional charge-density-wave material 1$T$-TaS$_2$, we
demonstrate that this type of pressure cell allows the determination
of the absolute value of the reflectivity even in the far-infrared
region. Furthermore, its precision is comparable to that of standard
ambient-condition reflectivity experiments.
\begin{figure}[h!]
\includegraphics[width=2.15in]{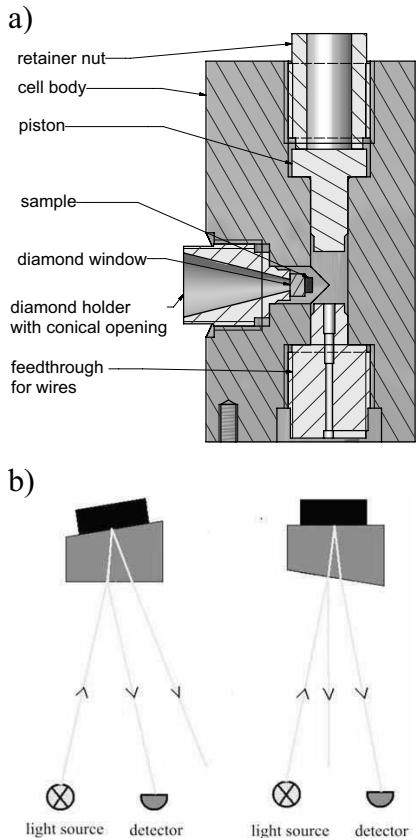}
\caption{(Color online) Our home-designed optical pressure cell:
steel body equipped with a type IIA diamond window. The upper panel
shows the prorata drawing of the cell. The sample is mounted on the
inner surface of the diamond window which has $d=1.5$\,mm free
aperture. As demonstrated in the lower panel, the angle of wedging
between the two surfaces of the window allows for a clean separation
of the reflections from the vacuum-diamond and diamond-sample
interface and thus facilitates reference measurements (see text for
details). \label{fig1}}
\end{figure}
Our design for the diamond-window based self-clamping type pressure
cell is shown in Fig.~\ref{fig1}a. The light access is achieved
through a type IIA diamond window fixed at the lateral part of the
metallic body with the beam being perpendicular to the horizontal
axis of the cell. In order to determine the absolute value of
reflectivity for the diamond-sample interface, $R_{ds}$, the window
has wedged plane surfaces which facilitates the measurement of
reference signal. It is also sufficient to eliminate interference
fringes coming from multiple reflections. Besides the optical
access, an electrical leadthrough is also implemented, which
facilitates either the in situ monitoring of pressure by a resistive
sensor (InSb, amorphous carbon, etc.) or the simultaneous
measurement of the sample resistivity during the pressure runs. For
low-temperature experiments the pressure cell is attached to the
cold finger of a the He flow cryostat and the thermalization is
ensured by a radiation shield.

As followed in the sketch of Fig.~\ref{fig1}b, the intensity
reflected back from the vacuum-diamond and diamond-sample interfaces
($I_{vd}$ and $I_{ds}$, respectively) can be detected separately by
a few degree (typically $\sim7^{\circ}$) rotation of the cell around
its vertical axis due to the wedging of the diamond. (A wedging
angle of $2^{\circ}$ results in $\sim10^{\circ}$ angular deviation
between the two reflected beams which, together with the finite beam
size and the rotation of the cell, requires an opening angle of
$\sim28^{\circ}$ for the conical bore on the diamond holder in our
optical arrangement.) The nearly normal incidence is still held for
the both positions. Correspondingly, the reflectivity of the sample
relative to the diamond can be obtained from the measured
intensities:
\begin{equation}
R_{ds}(\omega)=\frac{R_{vd}(\omega)}{(1-R_{vd}(\omega))^2}\cdot\frac{I_{ds}(\omega)}{I_{vd}(\omega)}\,
,
\label{eq1}
\end{equation}
where $R_{vd}$ ($\equiv R_d$) is the absolute reflectivity of the
diamond. The $R_{vd}(\omega)/(1-R_{vd}(\omega))^2$ prefactor can be
calculated from the well-documented refractive index of diamond,
$\hat{n}_d$.
\cite{Philipp1964,Edwards1981,Edwards1985,Djurisic1998,Thomas1995}
High-quality, such as type IIA, optical diamonds show no significant
absorption up to $\omega \approx 40000$\,cm$^{-1}$. On the other
hand, the effect of weak absorption introduces only another
frequency-dependent factor into Eq.~\ref{eq1} (which can be directly
determined from the transmittance of the window) but its influence
on $R_{vd}$ and $R_{ds}$ can be usually neglected. If
$R_{ds}(\omega)$ is measured over a sufficiently broad range of
energy, the complex refractive index of the sample relative to the
diamond $\hat{n}_{ds}(\omega)$ can be obtained by the Kramers-Kronig
analysis. Since $\hat{n}_d(\omega)$ is purely real and shows a
monotonous increase of about $10\%$ up to $\omega \approx
40000$\,cm$^{-1}$, it is straightforward to calculate the complex
dielectric response of the sample relative to the vacuum. This
analysis fails only for the multiphonon absorption bands of the
diamond located in the range of
$\omega=1500-2700$\,cm$^{-1}$.\cite{Thomas1995} We note at this
point that, additional to the absorption of the diamond above
$\omega \approx 40000$\,cm$^{-1}$, the high-energy limit of this
method is determined by the roughness and/or planarity of the
diamond-sample interface $\delta_{ds}$. Therefore, special care
should be taken for the proper matching between the window and the
sample in order to eliminate interference and diffraction effects
inherently appearing for wavelength shorter than $\delta_{ds}$.
\begin{figure}[th!]
\includegraphics[width=2.7in]{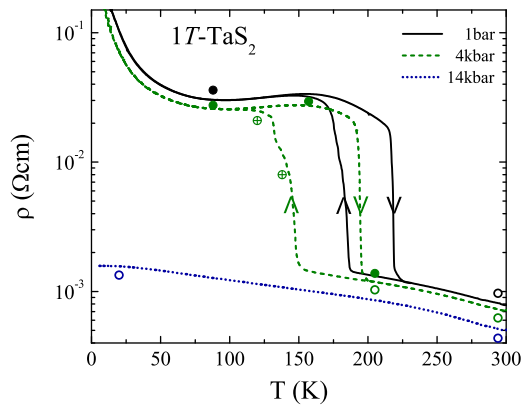}
\caption{(Color online) Temperature dependence of the resistivity
for three selected pressures in 1$T$-TaS$_2$. At each pressure
infrared reflectivity was measured at temperatures labeled by
different symbols: $\circ$, $\bullet$, and $\oplus$ correspond to
points in the pressure-temperature phase diagram which are
characterized by the reflectivity spectrum as the room-temperature,
low-temperature phase, or as a mixed state, respectively. At
$p=1$\,bar/$14$\,kbar the corresponding symbols are shown
above/below the resistivity curve. For $p=4$\,kbar the reflectivity
measurements performed in cooling run are indicated below while
those in warming up are shown above the $\rho(T)$ curve.}
\label{fig2}
\end{figure}
\begin{figure*}[th!]
\includegraphics[width=6.1in]{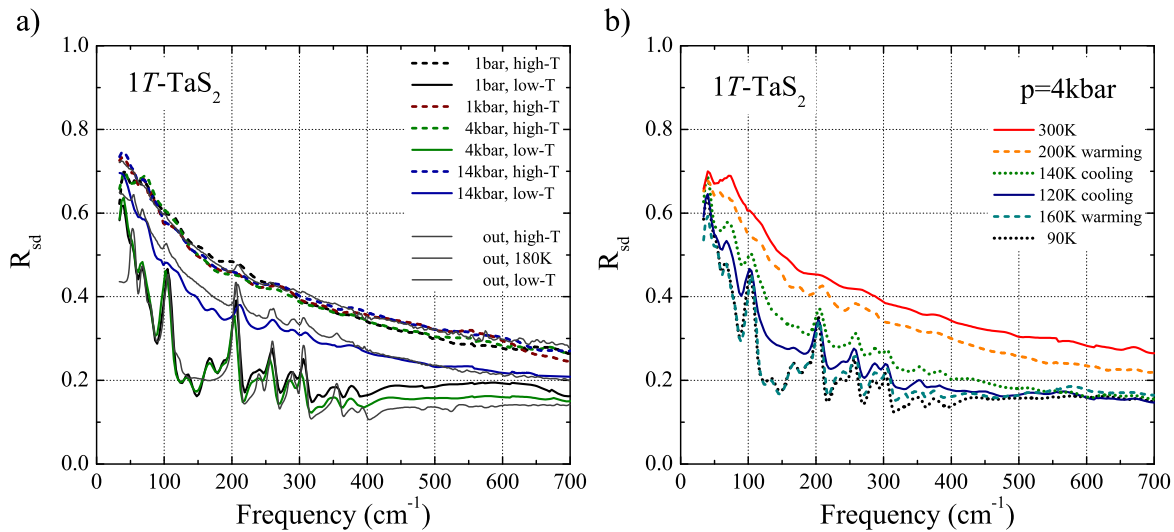}
\caption{(Color online) Far-infrared reflectivity of 1$T$-TaS$_2$
relative to diamond, $R_{ds}(\omega)$, over a broad range of
hydrostatic pressure and temperature. Left panel: reflectivity
spectra measured in the room-temperature and the low-temperature
phase of 1$T$-TaS$_2$ at various pressures. Temperatures
corresponding to the low-temperature measurements can be followed in
Fig.~\ref{fig2}. For the verification of the results obtained within
the pressure cell, $R_{ds}(\omega)$ spectra calculated from the
ambient-pressure absolute reflectivity (reproduced from
Ref.~\onlinecite{Gasparov2002}) are also plotted. Right panel:
far-infrared reflectivity of 1$T$-TaS$_2$ below room temperature at
$p=4$\,kbar.} \label{fig3}
\end{figure*}

In the following we shall present the high-pressure optical study of
the 2-dimensional charge-density-wave (CDW) system 1$T$-TaS$_2$
carried out by the above described pressure cell. At ambient
pressure and room temperature 1$T$-TaS$_2$ is a bad metal or
semi-metal characterized by a high and almost temperature
independent resistivity ($\rho\approx 10^{-3}$\,$\Omega$cm) and by a
fully incoherent low-energy optical conductivity.\cite{Gasparov2002}
With lowering the temperature, the material goes through a
first-order phase transition at $T_{lock}=180$\,K which is clearly
manifested in the electronic transport as it is accompanied by a
jump-like enhancement of the resistivity \cite{Tani1977} and by the
opening of a soft gap $\Delta\approx 800$\,cm$^{-1}$ in the optical
conductivity.\cite{Gasparov2002} Recent photoemission experiments
also reflect that the electronic structural changes are dominantly
restricted to the low-energy
region.\cite{Zwick1998,Perfetti2005,Perfetti2006,Clerc2006} On the
other hand, due to its 2-dimensional CDW ordering and the large
electron-phonon coupling,\cite{Perfetti2006,Clerc2006} the material
shows extremely high sensitivity to external
pressure.\cite{Tani1977} Therefore, 1$T$-TaS$_2$ offers an ideal
playground to study the pressure-induced electronic and structural
changes by infrared spectroscopy.

Prior to the optical experiments under pressure, we have
investigated the pressure-temperature phase diagram of the material
by means of dc resistivity. The results are shown in Fig.~\ref{fig2}
for three selected pressures $p=1$\,bar, $4$\,kbar, and $14$\,kbar.
In agreement with early data,\cite{Tani1977} the low-temperature
insulating phase is strongly suppressed although the resistivity is
almost unchanged in either of the phases. Already at $p=4$\,kbar the
transition is shifted to $T_{lock}\approx150$\,K and the hysteresis
is significantly broadened,\cite{note1} while at $p=14$\,kbar the
high-temperature phase is extended to the lowest temperatures. This
semi-metallic state exhibit nearly temperature independent charge
transport as the resistivity increases only by a factor of $\sim3$
below $T=300$\,K. We note here that the resistivity in the
insulating phase also shows anomalous behavior which may be
attributed to Mott-Anderson localization effects together with the
opening of the pseudo-gap $\Delta\approx 800$\,cm$^{-1}$.

The far-infrared reflectivity of 1$T$-TaS$_2$ was investigated below
room temperature for various pressures $p=1$\, bar, $1$\,kbar,
$4$\,kbar, and $14$\,kbar in the photon energy range of
$\omega=30-700$\,cm$^{-1}$ as shown in Fig.~\ref{fig3}. For
verification of the results, besides the $R_{ds}(\omega)$ spectra
measured inside of the pressure cell, Fig.~\ref{fig3}a also displays
spectra evaluated from the ambient pressure absolute reflectivity
$R_s(\omega)$ reproduced from Ref.~\onlinecite{Gasparov2002}. (For
these data the complex refractive index $\hat{n}_s$ was obtained by
Kramers-Kronig transformation and $R_{ds}$ was calculated with use
of the Fresnel equation from $\hat{n}_s$ and $\hat{n}_d$.) The
ambient pressure data measured inside and out of the pressure cell
show fairly good agreement both at room and low temperature
verifying the applicability of our method for the fully quantitative
measurement of reflectivity under pressure even in the hardly
accessible far-infrared spectral range.

Similarly to the resistivity, the room-temperature reflectivity of
the sample does not show significant variation up to $p=14$\,kbar.
The reflectivity spectrum in the low-temperature insulating phase is
reduced and exhibits sharp phonon resonances for
$\omega<400$\,cm$^{-1}$. Its variation with pressure is also
negligible for $p=1$\,bar$-4$\,kbar and probably the same remains
valid, as long as the insulating phase is not completely suppressed.
On the other, at $p=14$\,kbar when the semi-metallic phase becomes
the ground state, the low-temperature reflectivity is reminiscent to
that of at room temperature. Although it is somewhat reduced in
absolute value the spectral shape is still semi-metallic-like and
the phonon modes remain almost completely screened. For comparison,
we also plotted the reflectivity measured out of the cell just above
$T_{lock}=180$\,K. As expected from the tiny variation of the
resistivity at $p=14$\,kbar below room temperature, the two spectra
are close to each other over the whole far-infrared range.

The temperature dependence of $R_{ds}(\omega)$ has been studied in
more details, especially in the hysteretic region of the transition,
for the intermediate pressure $p=4$\,kbar (see Fig.~\ref{fig3}b).
The spectra follow the tendency of the $R(T)$ curve both in the
cooling and warming run as depicted by symbols plotted
simultaneously with the resistivity curves. Except for $T=140$ and
$120$\,K in cooling down, where the coexistence of the two phases
was observed in the reflectivity, $R_{ds}(\omega)$ spectra purely
showed either the semi-metallic or the insulating character.

In conclusion, we have extended the range of the high-pressure
optical spectroscopy to the far-infrared region keeping the accuracy
of ambient-pressure experiments. For this purpose we have developed
an optical pressure cell equipped with a wedged diamond window of
large free aperture by which reflectivity spectra can be measured
down to $\omega\approx20-30$\,cm$^{-1}$ for hydrostatic pressures up
to $p=26$\,kbar. This technique, which is strongly based on the
excellent optical properties of high-quality (type IIA) diamonds, is
well-suited to study the systematics of bandwidth-controlled
metal-insulator transitions in interacting electron systems via the
pressure-induced changes in their low-energy electronic structure.
Its applicability was demonstrated by the optical study of the
2-dimensional charge-density-wave 1$T$-TaS$_2$. We have pointed out
that the room-temperature semi-metallic phase can be effectively
extended by external pressure and stabilized as the ground state
above $p=14$\,kbar. Furthermore, the corresponding fully incoherent
low-energy optical conductivity is almost temperature independent
below $T=300$\,K. For intermediate pressures, the onset of the
low-temperature insulating phase is reflected by the sudden drop of
the reflectivity and by the emergence of sharp phonon resonances.

\section*{Acknowledgement}
The authors are grateful to L. Mih\'aly for useful comments and
discussions. They are also greatly indebted to L.V. Gasparov for
sending recently published ambient-pressure reflectivity spectra.
This work was supported by the Hungarian Scientific Research Funds
OTKA under grant Nos. F61413 and K62441, by the Swiss NSF, and its
NCCR "MaNEP" and by the DOE under contract number DE-AC02-98CH10886.
I. K. was a grantee of Bolyai J\'anos Fellowship.

%
%

%

\begin{references}
%
\bibitem{Mao1990}H.K. Mao, R.J. Hemley, and M. Hanfland, \prl \textbf{65}, 484 (1990).
%
\bibitem{Hemley1996}R.J. Hemley, H.K. Mao, A.F. Goncharov, M. Hanfland, and V. Struzhkin, \prl \textbf{76}, 1667 (1996).
%
\bibitem{Sacchetti2006}A. Sacchetti, M.C. Guidi, E. Arcangeletti, A. Nucara, P. Calvani, M. Piccinini, A. Marcelli,
and P. Postorino, \prl \textbf{96}, 35503 (2006).
%
\bibitem{Garcia2004}J.L. Garcia-Munoz, M. Amboage, M. Hanfland, J.A. Alonso, M.J. Martinez-Lope, and R. Mortimer, \prb \textbf{69}, 94106 (2004).
%
\bibitem{Baldassarre2007}L. Baldassarre, A. Perucchi, E.
Arcangeletti, D. Nicoletti, D. Di Castro, P. Postorino, V.A.
Sidorov, and S. Lupi, \prb \textbf{75}, 245108 (2007).
%
\bibitem{Ohmura2006}A. Ohmura, A. Machida, T. Watanuki, and K. Aoki, \prb \textbf{73}, 104105 (2006).
%
\bibitem{Sacchetti2007}A. Sacchetti, E. Arcangeletti, A. Perucchi, L. Baldassare, P. Postorino, S. Lupi, N. Ru,
I.R. Fisher, and L. Degiorgi, \prl \textbf{96}, 35503 (2007).
%
\bibitem{Nishi2005}T. Nishi, T. Takahashi, Y. Mori, Y.S. Kwon, H.J. Im, and H. Kitazawa, \prb \textbf{71}, 220401 (2005).
%
\bibitem{Kuntscher2005}C.A. Kuntscher, S. Frank, I. Loa, K. Syassen, T. Yamauchi, and Y. Ueda, \prb \textbf{71}, 220502(R) (2005).
%
\bibitem{Sun2006}L. Sun, A.L. Ruoff, C.S. Zha, G. Stupian, J. Phys.
D \textbf{18}, 8573 (2006).
%
\bibitem{Philipp1964}H.R. Philipp and E.A. Taft, Phys. Rev. \textbf{136},
A1445 (1964).
%
\bibitem{Edwards1981}D.F. Edwards and E. Ochoa, J. Opt. Soc. Am. \textbf{71}, 607
(1981).
%
\bibitem{Edwards1985}D.F. Edwards and H.R. Philipp, \textit{Handbook of Optical Constants of
Solids}, pp. 665-673, edited by E.D. Palik, Academic, Orlando,
Florida (1985).
%
\bibitem{Djurisic1998}A.B. Djurisic and E.H. Li, Appl. Optics \textbf{37}, 7273 (1998).
%
\bibitem{Thomas1995}M.E. Thomas, W.J. Tropf, and A. Szpak, Diamond
Films and Tech. \textbf{5}, 159 (1995) and references therein.
%
\bibitem{Tani1977}T. Tani, T. Osada, and S. Tanaka, Solid State
Commun. \textbf{22}, 269 (1977).
%
\bibitem{Gasparov2002}L.V. Gasparov, K.G. Brown, A.C. Wint, D.B. Tanner, H. Berger, G. Margaritondo, R. Gaal, and L. Forro, \prb \textbf{66}, 94301 (2002).
%
\bibitem{Zwick1998}F. Zwick, H. Berger, I. Vobornik, G. Margaritondo, L. Forro, C. Beeli, M. Onellion, G. Panaccione, A. Taleb-Ibrahimi, and M. Griono, \prl \textbf{81}, 1058 (1998).
%
\bibitem{Perfetti2005}L. Perfetti, T.A. Gloor, F. Mila, H. Berger, and M. Grioni, \prb \textbf{71}, 153101 (2005).
%
\bibitem{Perfetti2006}L. Perfetti, P.A. Loukakos, M. Lisowski, U. Bovensiepen, H. Berger, S. Biermann, P.S. Cornaglia, A. Georges, and M. Wolf, \prl \textbf{97}, 67402 (2006).
%
\bibitem{Clerc2006}F. Clerc, C. Battaglia, M. Bovet, L. Despont, C. Monney, H. Cercellier, M.G. Garnier, P. Aebi, H. Berger, and L. Forro, \prb \textbf{74}, 155114 (2006).
%
\bibitem{note1}In our convention $T_{lock}$ is defined by the
temperature the onset of the transition in cooling down.
\end{references}
\end{document}